\documentclass[epj]{svjour}
 \usepackage{graphics}
\usepackage{psfig}% etc
\begin{document}
\title{The reaction {\boldmath$\pi N \rightarrow \pi \pi N$}
 in a meson-exchange approach}
\author{S. Schneider\inst{1} 
\thanks{\emph{Present address:} Carl Zeiss SMT AG, Oberkochen,
Germany.}%
\and S. Krewald\inst{1}
\and
Ulf-G. Mei{\ss}ner\inst{1},\inst{2}
}                     % Do not remove
\institute{
Institut f\"ur Kernphysik, Forschungszentrum 
J\"ulich,  D-52425 J\"ulich, Germany
 \and
Universit\"at Bonn, Helmholtz-Institut f\"ur Strahlen- 
und Kernphysik,  Nu{\ss}allee 14-16, D-53115 Bonn, Germany 
}
%\preprint{FZJ-IKP-TH-2006-8, HISKP-TH-06/08}
%
\date{Received: date / Revised version: date}
\abstract{
A  resonance  model for two-pion production in the 
pion-nucleon reaction is developed that includes  information
obtained in the analysis of pion-nucleon scattering in a meson-exchange model. 
The baryonic resonances $\Delta(1232)$, $N^*(1440)$, $N^*(1520)$,  
$N^*(1535)$, and $N^*(1650)$ are included.
The model  reproduces the total cross sections up to
kinetic energies of the incident pion of 350~MeV  and obtains the
shapes of the differential cross sections in reasonable agreement with the data.
 % end of PACS codes
} %end of abstract
\maketitle
\section{Introduction}
\label{intro}
Recent experimental progress provides data for two-pion production
in both pion-induced and electromagnetic reactions up to energies of about 
2.1~GeV, see e.g. Refs.~
\cite{Fatemi:2003yh,Langgartner:2001sg,Wolf:2000qt,Prakhov:2004zv,Kermani:1998gp}.
New isobar models have been developed to deduce  masses and widths 
of baryon resonances from the data~\cite{Aznauryan:2005tp,Anisovich:2004zz}.
 A basic problem for these analyses is given by the
fact that the excitation of resonances is accompanied by other non-resonant
processes, the so-called background. At present, the non-resonant processes
are treated phenomenologically~\cite{Aznauryan:2005tp},
 using or extending  methods known from the
analysis of pion-nucleon scattering~
\cite{Vrana:1999nt,Arndt:2002xv,Tiator:2003uu,Feuster:1997pq}.
 A combined 
theoretical treatment of resonances and background is a challenge for 
theory. 

At low energies, chiral perturbation theory provides a quantitative theoretical 
understanding of pion-induced two-pion production.
 Calculations in heavy baryon chiral
perturbation theory have been extended to third 
order~\cite{Bernard:1995gx,Fettes:1999wp,Mobed:2005av}.
An important result obtained in Ref.~\cite{Fettes:1999wp} is the observation
that contributions from
loop diagrams are negligible. This is non-trivial because in pion-nucleon
scattering, unitarity effects can be important even close to threshold in some
partial waves. In two-pion production, however, the imaginary contributions interfere
destructively. The important contributions at third order are  the
tree level diagrams involving the finite dimension two low--energy 
constants $c_i$ and tree level
corrections of order $1/m_N$, with $m_N$ the nucleon mass. 
This finding of Ref.~\cite{Fettes:1999wp} explains why a previous relativistic
baryon chiral perturbation theory calculation at tree level including terms
from the dimension two effective pion-nucleon Lagrangian works so well for
pion kinetic energies up to 400 MeV~\cite{Bernard:1997tq}.
Moreover, the numerical values of the low--energy constants (LECs) 
$c_i$ of the pion-nucleon
Lagrangian can be understood in terms of  resonance saturation~\cite{Bernard:1996gq},
more precisely through the $s$- and $u$-channel excitations of baryon
resonances ($\Delta, N^* (1440)$) and $t$-channel meson resonances
($\sigma, \rho$).  
As a consequence, one may obtain a reasonable model for treating baryon
resonances in two-pion production reactions by replacing the low--energy
constants of chiral perturbation theory by diagrams containing explicit
resonances. This has been done by Jensen and Miranda~\cite{Jensen:1997em}
and by Kammano and Arima~\cite{Kammano:2004mx} and  in resonance models like 
e.g.~\cite{Oset:1985wt,Jaekel:1992nz,GomezTejedor:1995pe,Nacher:2000eq}.
 Such an approach makes sense if one checks that at low
energies the chiral perturbation theory results are recovered or if one enforces
this behaviour through explicit matching of the pertinent amplitudes.
% Recently,
%a chiral unitary approach for meson-baryon scattering has been used to
%study photon-induced $\pi^0\eta$ and $\pi^0K^0$ production
%\cite{Doring:2005bx}.

In the present work, we want to include information about pion-nucleon
 scattering into the construction of a resonance model
for two-pion production. We use results obtained in Ref.~\cite{Krehl:1999km}
which treats pion-nucleon scattering
in a coupled-channel model. The model incorporates
 the effects of the $\pi\pi N$ states by introducing $\sigma N$, $\pi \Delta$
and $\rho N$ channels.  The interactions between these channels is derived 
from an chirally symmetric Lagrangian 
supplemented by additional terms for the $\Delta$, $\omega$, $\rho$, and
$\sigma$ fields. The model includes the $\Delta(1232)$, $N^*(1520)$,
$N^*(1535)$, and $N^*(1650)$ as explicit resonances. It  is able to reproduce
the experimental phase shifts and inelasticities up to 1500 MeV. 
For larger energies, it predicts a non-resonant background.
The model does not need an explicit Roper resonance $N^*(1440)$, but is
able to describe the $P_{11}$ partial wave and in particular the 
inelasticity by the  dynamics of the $\sigma N$ and $\pi \Delta$
channels.  
 
\vfill

%%%%%%%%%%%%%%%%%%%%%%%%%%%%%%%%%%%%%%%%%%%%%%%%%%%%%%%%%%%%%%%%%%%%%%%%%%
\section{The model}
\label{sec:1}

\begin{figure}
\resizebox{0.5\textwidth}{!}{
  \includegraphics{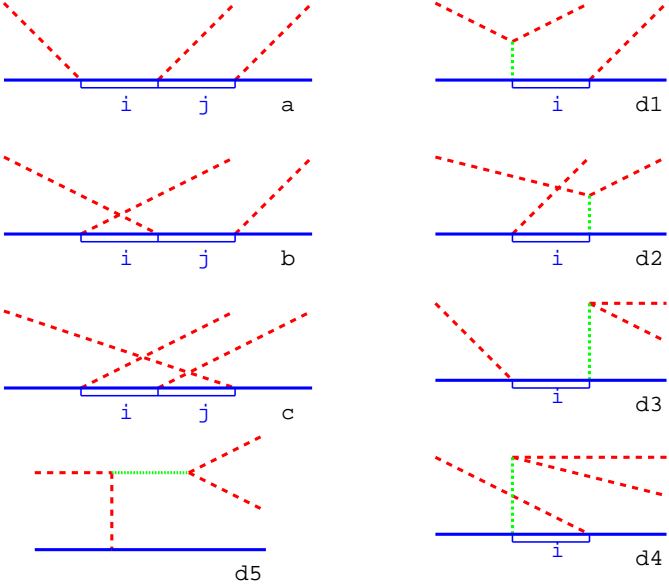}
}
\caption{
Classes of diagrams describing the $\pi N \to \pi\pi N$
 reaction near threshold. The internal baryon propagators labeled $i$ and $j$
(blue double lines)
represent the nucleon, the $\Delta$, and the $N^*$ resonances
$N^*(1440)$, $N^*(1520)$, $N^*(1535)$, $N^*(1650)$, respectively. 
The $\rho$ and $\sigma$ meson propagators
are shown by (green) dotted lines. The 
(red) dashed lines stand for external pions, while (blue)
solid lines are external nucleons.
}
\label{fig:a}       % Give a unique label
\end{figure}

Fig.~\ref{fig:a} 
shows the classes of diagrams which are included in the present resonance model. 
The diagrams (a), (b), and (c) contain two baryon propagators whereas (d1) -
(d4) contain one. We consider the nucleon, the $\Delta (1232)$,
 and the $N^*(1520)$, $N^*(1535)$, $N^*(1650)$, and $N^*(1440)$ which are denoted 
by the indices $i$ and $j$, respectively. We also include two--pion resonances
in the $I_{\pi\pi} =0$ and $I_{\pi\pi} =1$ partial waves, see diagram (d5)
(and below).

Meson-dynamical models generate unitary pion-nucleon scattering $T$-matrices
by solving the Bethe-Salpeter equation
\begin{equation}
T = K + K \, G \, T,
\label{eq:1}
\end{equation}
 where $G$ denotes the two-particle propagator and $K$ 
the scattering kernel which in principle
includes all two-particle irreducible diagrams.
The scattering kernel can be split into a pole term $K_2$ which contains
$s$-channel pole diagrams and the non-pole term $K_1 = K -K_2$.
As is well known, the $T$-matrix separates into a pole part $T_2$ and
a non-pole part $ T_1 = T - T_2$ \cite{Haymaker:1969id,Afnan:1980am}.
The non-pole $T$-matrix is obtained by solving 
\begin{equation}
T_1 = K_1 + K_1 G T_1~.
\label{eq:2}
\end{equation}
The non-pole $T$-matrix generates the dressed vertex $f$ which is obtained 
from the bare vertex $f_0$ as follows:
\begin{equation}
f = f_0 + f_0 \, G \, T_1~.
\label{eq:3}
\end{equation}
The self--energy $\Sigma$ is given by
\begin{equation}
\Sigma = f_0^\dagger \, G \, f~, 
\label{eq:4}
\end{equation}
and the pole $T$-matrix reads
\begin{equation}
T_2 = f^\dagger \, g \, f~, 
\label{eq:4a}
\end{equation}
where $ g^{-1} = G^{-1} - \Sigma$ denotes the dressed propagator.

The diagram (d1) shown in Fig.~\ref{fig:a} factorizes into two subdiagrams, the
first of which is a 
part of the non-pole scattering kernel of the meson-nucleon $T$-matrix 
corresponding to
a  meson-exchange in  the $t$-channel, while the second subdiagram 
is the coupling of the intermediate baryon propagator to a
pion-nucleon final state.
Likewise, the diagram (d2) factorizes into a pion production followed 
by a subdiagram corresponding to a non-pole meson-baryon scattering kernel.
Diagrams (b) and (d) factorize into $u$-channel processes and pion production
amplitudes. The $u$-channel  and $t$-channel subdiagrams
 are  the Born approximation
to the non-pole $T$-matrix.

In the present work, we restrict the model to the diagrams shown in
 Fig.~\ref{fig:a}. This allows a relatively simple treatment of two-pion
production because only tree-level diagrams with dressed vertices and
dressed propagators taken from a pion-nucleon scattering model 
have to be evaluated.  An iteration of the
non-pole subdiagrams would require to solve Eq.~(\ref{eq:2}) which 
commonly is done numerically relying on a partial wave representation
in the center-of-mass frame, but implies technical complications
when boosts to other frames are required, e.g. when evaluating the
two-pion production amplitude.

\renewcommand{\arraystretch}{1.3}
\begin{table}
\caption{The meson-baryon interaction Lagrangian.}
\label{tab:1}
\begin{center}
\begin{tabular}{|l|l|}
\hline%\noalign{\smallskip}
vertex & $\qquad \qquad\qquad\cal{L}$\\
\hline%\noalign{\smallskip}
$\pi N N $   &
 $-\frac {f_{\pi N N}} {M_\pi} \bar{\psi}\gamma_5 \gamma_\mu \tau_a \partial^\mu \pi_a \psi $\\
$\pi N \Delta $   &
 $\frac {f_{\pi N \Delta}} {M_\pi} \bar{\Delta}^\mu  T_a^{\dagger} \partial^\mu \pi_a \psi $ + h.c.\\
$\rho \pi \pi $   &
 $- g_{\rho \pi \pi}  \epsilon_{abc}  \pi_a \partial^\mu \pi_b \rho_c$ \\
$\rho N N $   &
 $-g_{\rho N N}  \bar{\psi} ( \gamma_\mu - \frac {\kappa_{\rho N N}}{2 m_N}  
    \sigma^{\mu\nu} \partial_\nu ) 
    \tau_a  \rho^a_\mu \psi $\\
$\sigma \pi \pi $   &
$ - g_1 M_\pi^2 \pi_a \pi_a \sigma + \frac {g_2}{2} \partial_\mu \pi_a \partial^\mu \pi_a \sigma $\\
$\sigma N N $   &
$- g_{\sigma N N} \bar{\psi} \psi \sigma $   \\
$\sigma \sigma \sigma $   &
$- g_{\sigma \sigma \sigma } M_\sigma \, \sigma^3 $   \\
$\pi \Delta \Delta $   &
 $\frac {f_{\pi \Delta \Delta}} {M_\pi} \bar{\Delta}_\mu 
 \gamma_5 \gamma_\nu
 T_a \partial^\nu \pi_a \Delta^\mu $ \\
$\rho N \Delta $   &
 $ - i \frac {f_{\rho N \Delta}} {M_\rho} \bar{\Delta}^\mu 
 \gamma_5 \gamma^\nu T_a^\dagger \rho^a_{\mu\nu} \psi + h.c.$ \\
$\rho \Delta \Delta $   &
 $-g_{\rho N N}  \bar{\Delta}_\sigma
 ( \gamma_\mu - \frac {\kappa_{\rho \Delta \Delta}}{2 m_\Delta}  
    \sigma^{\mu\nu} \partial_\nu ) 
T_a \rho^a_\mu \Delta^\sigma $ \\
$ P_{11}\pi N $   &
$- \frac{f_{P11\pi N}} {M_\pi} \bar{\psi}_{N^*}
 \gamma_5 \gamma_\mu \tau_a \partial^\mu \pi_a \psi  + h.c. $\\
$ P_{11}\pi \Delta $   &
$  \frac{f_{P11\pi \Delta}} {M_\pi} 
  \bar{\Delta}^\mu  T_a^\dagger 
  \partial_\mu \pi_a \psi_{N^*}  + h.c. $\\
$ P_{11}\sigma \Delta $   &
$- g_{P_{11}\sigma N } \bar{\psi}_{N^*}  \psi \sigma  + h.c.$   \\
$ D_{13}\pi N $   &
$i \frac{f_{D13\pi N}} {M_\pi^2} \bar{\psi}_{N^*}^\mu
 \gamma_5 \gamma^\nu \tau_a \psi \partial_\nu \partial_\mu \pi_a + h.c. $\\
$ D_{13}\pi \Delta $   &
$  \frac{f_{D13\pi \Delta}} {M_\pi} \bar{\psi}_{N^*}^\mu
\gamma_\nu T_a \partial^\nu \pi_a \Delta_\mu + h.c. $\\
$ S_{11}\pi N $   &
$  \frac{f_{S11\pi N}} {M_\pi} \bar{\psi}_{N^*} \gamma_\mu \tau_a \partial^\mu \pi_a \psi + h.c.$\\
$ S_{11}\eta N $   &
$  \frac{f_{S11\eta N}} {M_\eta} \bar{\psi}_{N^*} \gamma_\mu  \partial^\mu \eta \psi + h.c.$\\
%\noalign{\smallskip}
\hline
\end{tabular}
\end{center}
\end{table}

In Ref.~\cite{Krehl:1999km},
 the inelasticities of the $P_{11}$ partial wave
were explained by the final--state interactions, i.e. the non-pole
contribution of the $T$-matrix.
In the present model, such a structure cannot be generated.
We therefore have  to treat the Roper in a simplified way and represent it
as an $s$-channel resonance. A Roper propagator is introduced which has a
 structure analogous to the nucleon propagator.

We  simplify the formalism employed in Ref.~\cite{Krehl:1999km}
 by solving the coupled channel problem for meson-nucleon scattering in
the $K$-matrix approximation. Moreover, we employ a derivative coupling for the
coupling of the nucleon resonances to the pseudo-scalar mesons which
was used  in Ref.~\cite{Gasparyan:2003fp} to improve the description of
the $S_{11} \pi N$ partial wave. 

\renewcommand{\arraystretch}{1.2}
\begin{table}
\caption{ Coupling constants and masses in GeV. 
The constants $ g_1^2$ and $g_2^2$ are in GeV$^{-2}$.}
\label{tab:2}       % Give a unique label
\begin{center}
\begin{tabular}{|l|l||l|l|}
\hline%\noalign{\smallskip}
$f^2_{ \pi NN }$/(4$\pi$)       & 0.0778     & $f^2_{P11(1440) \pi N}$/(4$\pi$)   & 0.011 \\
$f^2_{ \pi N \Delta }$/(4$\pi$) & 0.36       & $f^2_{P11(1440) \pi \Delta}$/(4$\pi$) & 0.04 \\
$g^2_{ \rho NN }$/(4$\pi$)       & 0.80     & $f^2_{P11(1440) \sigma N}$/(4$\pi$) & 13.0 \\
$\kappa_{ \rho NN }$        & 1.94     & $f^2_{D13(1520) \pi N}$/(4$\pi$) & 0.0009 \\
$g^2_{ \sigma NN }$/(4$\pi$)       & 1.03     & $f^2_{D13(1520) \pi \Delta}$/(4$\pi$) & 0.03 \\
$f^2_{ \pi \Delta \Delta }$/(4$\pi$) & 0.04       & $f^2_{S11(1535) \pi N}$/(4$\pi$) & 0.003 \\
$f^2_{ \rho N \Delta }$/(4$\pi$) & 4.5       & $f^2_{S11(1535) \eta N}$/(4$\pi$) & 0.47 \\
$g^2_{ \rho \Delta \Delta }$/(4$\pi$) & 16.0      & $f^2_{S11(1650) \eta
  N}$/(4$\pi$) & 0.009 \\
$\kappa_{ \rho \Delta \Delta }$        & 15.0     & &  \\
%\noalign{\smallskip}
\hline
$g^2_{ \rho \pi \pi }$/(4$\pi$)       & 2.905     & $g^2_{1}$/(4$\pi$) & 98.94 \\
$g^2_{ \sigma \sigma \sigma }$/(4$\pi$)       & 0.625     & $g^2_{2}$/(4$\pi$) & 7.32 \\
%\noalign{\smallskip}
\hline
$m_{ N }$      & 0.93893 &   $m_{D13(1520)}$  & 1.515 \\
$m_{ \Delta }$      & 1.232 & $m_{S11(1535)}$ & 1.535 \\
 $m_{P11(1440)}$ &  1.491 & $m_{S11(1650)}$ & 1.701 \\ 
%\noalign{\smallskip}
\hline
$M_{ \pi }$      & 0.13803 &   $M_{\rho}$  & 0.772 \\
$M_{ \eta }$      & 0.5473 &   $M_{\sigma}$  & 0.8346 \\
%\noalign{\smallskip}
\hline
\end{tabular}
\end{center}
\end{table}

For brevity, we only display the various propagators that enter the
calculation. All the necessary formalism to calculate the relevant amplitudes,
cross sections etc. can e.g. taken from Ref.~\cite{Fettes:1999wp}.

The nucleon propagator is given by:
\begin{equation}
 S_{N}(p) = \frac { \slash\hspace{-0.18cm}p + m_N}{p^2 -m_N^2 - \Sigma_N(p^2)}~,
\label{eq:6}
\end{equation}
with $m_N$ the nucleon mass. The self energy $\Sigma_N$ is obtained
by including the $\pi N$, $\pi \Delta$, and the $\sigma N$ reaction
channels in  Eq.~(\ref{eq:3}).  
The propagator of the $\Delta (1232)$ is given by
\begin{equation}
 D^{\mu\nu}(p) = \frac { \slash\hspace{-0.18cm}p + m_{\Delta}}
{p^2 -m_{\Delta}^2 - \Sigma_{\Delta}(p^2)}(P^{\frac {3}{2}})^{\mu\nu}
+  D^{\mu\nu}_{\frac {1}{2}}~,
\end{equation}
with $m_\Delta$ the mass of the  $\Delta (1232)$.
The spin-1/2 contribution $ D^{\mu\nu}_{\frac {1}{2}} $ is identical with
the spin-1/2 part of the Rarita-Schwinger propagator. These contributions
are, however, non--propagating and can be represented in an effective field
theory approach by contact operators, see e.g.~\cite{Bernard:2003xf}.
The delta self--energy $\Sigma_\Delta$ is evaluated taking into account
only the pion-nucleon intermediate state and neglecting the non-pole
contribution of the pion-nucleon interaction, i.e.:
\begin{equation}
\Sigma_{\Delta} = f_0^\dagger \, G_{\pi N} \, f_0~. 
\end{equation}
This is a good approximation because the $P_{33}$ partial wave in the
pion-nucleon scattering model of Ref.~\cite{Krehl:1999km}
 is given
mainly by the $\Delta$-pole diagram for partial waves up to
approximately 1.3 GeV.

The interaction Lagrangians employed are shown in Table~\ref{tab:1}.
A partial refit of the parameters of Ref.~\cite{Krehl:1999km} is required.
In a first step, the parameters of the $\pi N$ potential are readjusted 
without the $N^*$ pole diagrams to the scattering lengths and phase shifts below
1.2 GeV. Then the $\pi \Delta \Delta, \rho N \Delta $, and the
$\rho \Delta \Delta$ couplings are fitted.
Finally the parameters of the $N^*$ resonances are
fixed. Table~\ref{tab:2} summarizes the coupling constants and masses.
The value of the tensorial $\rho$NN
vertex has been reduced in comparison to the value employed in 
Ref.~\cite{Krehl:1999km}.

The diagrams (d1), (d2), (d3), (d4),  and (d5) contain a $\sigma$ or $\rho$
propagator. Since here we need a reasonable model for the
pion-pion interaction below two-pion invariant masses of 1~GeV only, 
a simplified version of the meson-exchange model to pion-pion scattering suffices
\cite{Sassen:2002qv}.
 We include a dressing of the $\sigma$ and $\rho$ propagators 
by two-pion intermediate states only, neglecting $K\bar{K}$ intermediate states
and non-pole scattering kernels. 
Tables~\ref{tab:1} and~\ref{tab:2} show the relevant interaction
Lagrangian and coupling constants.
Unitarity is taken care of by solving 
Eq.~(\ref{eq:1}) for the two-pion system.

\section{Results}

\begin{figure}
\resizebox{0.50\textwidth}{!}{
\includegraphics{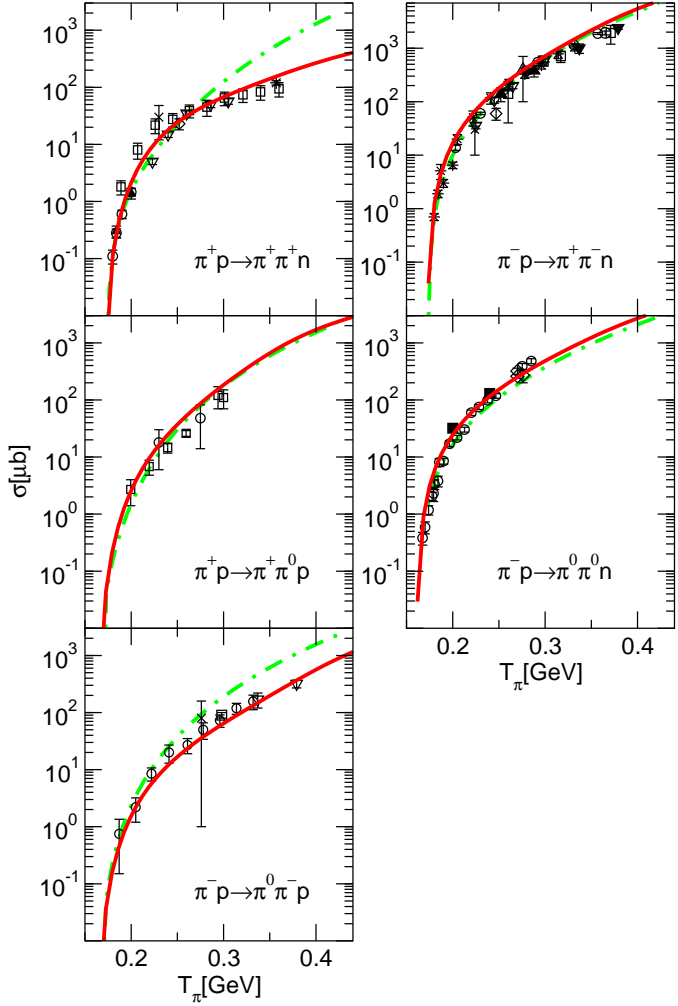}
}
\caption{Total cross sections for the reaction $  \pi N \to
\pi \pi N$. The (red) solid lines show the results of the present model. For 
comparison, the results obtained in Heavy Baryon chiral perturbation
theory~\cite{Fettes:1999wp}  
are displayed by the (green) dot-dashed line. The data are taken from
Ref.~\cite{Kermani:1998gp}
 and the compilation in Ref.~\cite{Vereshagin:1995mm}.
}
\label{fig:1}       % Give a unique label
\end{figure}

In Fig.~\ref{fig:1}, 
 the threshold behaviour  of the total cross sections
of the five different experimentally accessible channels
\begin{eqnarray} 
\pi^{\pm}\, p &\rightarrow& \pi^{\pm}\, \pi^+\, n~,\nonumber\\
\pi^{\pm}\, p &\rightarrow& \pi^{\pm}\, \pi^0\, p~,\nonumber\\
\pi^{-}\, p &\rightarrow& \pi^{0}\,\pi^0\, n~,
\end{eqnarray} 
are shown as a function of the kinetic energy of the initial state pion in
the laboratory frame, $T_\pi$. The corresponding invariant mass $\sqrt{s}$ of
the initial pion-nucleon system is given by 
\begin{equation}
s = (m_N + M_\pi)^2 + 2 m_N T_\pi~,
\end{equation}
with $M_\pi$ the relevant neutral or charged pion mass depending on the
channel under consideration.
The $\Delta$--isobar corresponds to $T_\pi = 0.19\,$GeV, while the maximal
pion kinetic energy $T_\pi = 0.4$ GeV translates into $\sqrt{s}=1.38\,$GeV which
is below the nominal mass of the Roper resonance. 
The observables are calculated assuming isospin symmetry. Following 
Refs.~\cite{Fettes:1999wp,Bernard:1997tq},
we included the effect of isospin breaking by the masses of the final states by
shifting the isospin symmetric threshold to the correct threshold energy
for each reaction channel. 

Close to threshold, the present model obtains cross sections which agree with the
results of chiral perturbation theory. This could be expected since the model
incorporates the experimental information on low energy pion-nucleon 
scattering. It should be stressed, however, that the model is not as precise as
chiral perturbation theory because it does not offer a
counting scheme which would allow systematic improvements.
Chiral perturbation theory starts to overshoot the experimental
cross sections for  the $\pi^{+} p \rightarrow \pi^{+}\pi^+ n$ reaction
at about $T_\pi \simeq 0.3$ GeV. 
Increasing the
order of the expansion would push  the range of validity of 
the chiral calculation to higher energies. In the model, some higher order terms
are included by unitarization effects. It is important to note that these
higher order effects are of no relevance in the threshold region.
The model continues to agree with
the experimental data up to about $T_\pi \simeq 0.4$ GeV.

\begin{figure}
\resizebox{0.50\textwidth}{!}{
  \includegraphics{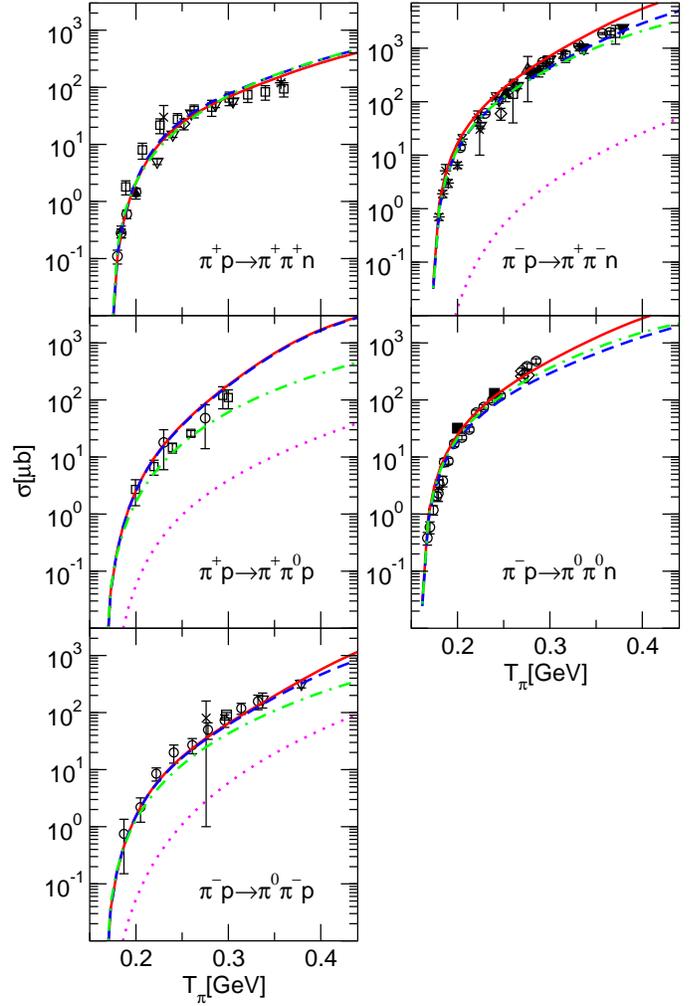}
}
\caption{Influence of resonances on the total cross sections
 for the reaction $  \pi N \to \pi \pi N$.
Full model: (red) solid lines; omission of the $\Delta$ resonance: 
(green) dot-dashed lines;
omission of the $N^*$ resonances: (blue) dashed lines. 
The (violet) dotted lines show the 
contribution of diagrams with an intermediate $\rho$-meson decaying to two pions.
 }
\label{fig:2}       % Give a unique label
\end{figure}

In Fig.~\ref{fig:2}, we investigate the effect of different resonances on the
total cross sections by omitting the contribution of various baryon resonances. 
The $\Delta$-resonance may be expected to play an
important role for the $\pi^+ p$ reaction for pion kinetic energies of about
$T_{\pi}=0.2$ GeV. This is seen in the reaction
 $ \pi^+ p \rightarrow \pi^+ \pi^0 p$,  but not in the reaction 
 $ \pi^+ p \rightarrow \pi^+ \pi^+ n$. 
A closer inspection showed that the contribution of a diagram with an 
intermediate $\Delta$ resonance by itself is large, but that there is destructive
interference between the $\Delta$ contributions. Such cancellations are also
observed in the description of electromagnetic two-pion production off
nucleons in the threshold region, see e.g.~\cite{Bernard:1994ds}.
The effect of the $\Delta$ can be  seen in the $\pi^- p$ reaction channels.
The impact of the $N^*$ resonances $S_{11}(1535)$ and $D_{13}(1520)$ on the
total cross sections is negligible in each of the five reaction channels in
the energy region investigated. The largest effect of $N^*$ resonances
is due to the Roper resonance and can be seen in the 
$ \pi^- p \rightarrow \pi^+ \pi^- n$ and
$ \pi^- p \rightarrow \pi^0 \pi^0 n$ reactions, where the two final pions
can be produced in a relative $s$-wave. The role of the Roper was recently also
investigated in Ref.~\cite{Kamano:2006vm}.
The $\sigma N$ channel is not allowed in
the  $ \pi^- p \rightarrow \pi^0 \pi^- p$ reaction and therefore the $N^*$
resonances  contribute  very little.

The  model  includes two-pion production via an intermediate rho-meson.
For the $\pi^+\pi^+n$ and $\pi^0\pi^0n$ final states, trivial isospin 
selection rules do not allow contributions from an intermediate rho-meson,
and in the other final states, the effect of those processes is marginal,
as expected, see Fig.~\ref{fig:2}.

\begin{figure}
\resizebox{0.50\textwidth}{!}{
  \includegraphics{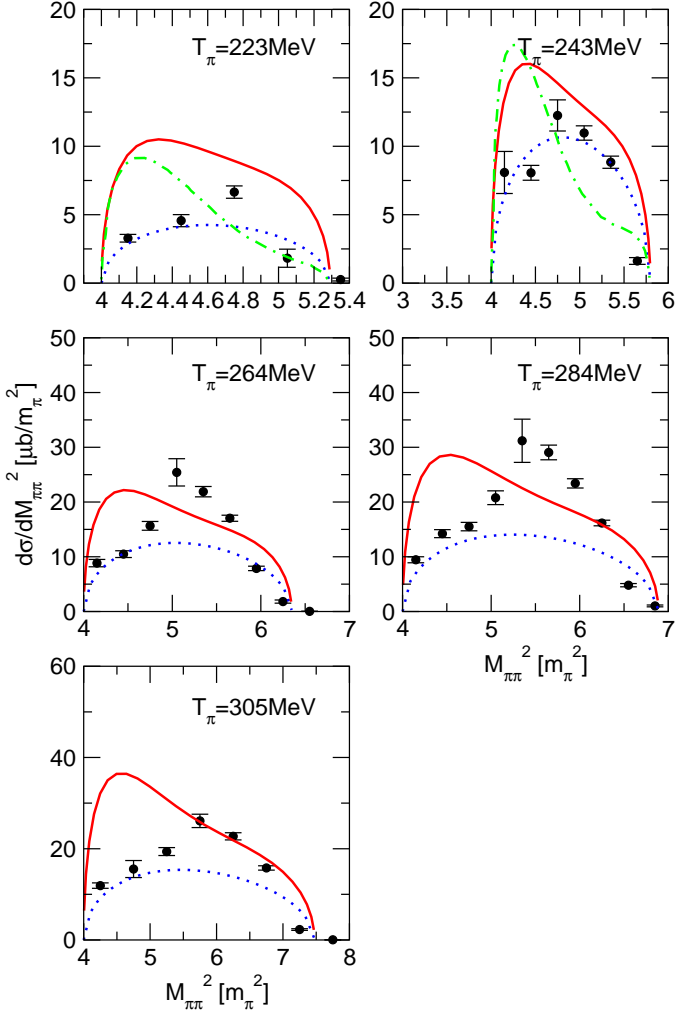}
}
\caption{Differential cross sections $d\sigma / dM^2_{\pi\pi}$
 for  $  \pi^+ p \rightarrow \pi^+ \pi^+ n$.
Full model: (red) solid line; phase space: (blue) dotted line; chiral perturbation theory:
(green) dot-dashed line~\cite{Fettes:1999wp}.
The data are from Ref.~\cite{Kermani:1998gp}.
        }
        \label{fig:3}       % Give a unique label
        \end{figure}

        The differential cross sections ${d\sigma}/{dM_{\pi\pi}^2}$ for the reactions 
        $  \pi^+ p \rightarrow \pi^+ \pi^+ n$ are shown for five different values
        of the kinetic energy $T_\pi$ of the incident pion in Fig.~\ref{fig:3} (here
        $M_{\pi\pi}$ is the invariant mass of the final-state two-pion system). 
        The data measured at
        TRIUMF can be fitted reasonably well by  three-body phase space~
\cite{Kermani:1998gp}.
Chiral perturbation theory (CHPT) predicts the average magnitude of the cross sections
correctly, but produces a large deviation from phase space emphasizing
low values of $M_{\pi\pi}^2$. The cross sections calculated within the 
resonance model agree with chiral
perturbation theory for low values of $M_{\pi\pi}^2$ for both $T_{\pi\pi}=223$ MeV
and $T_{\pi\pi}=243$ MeV, but  drop with increasing $M_{\pi\pi}^2$
more slowly than the CHPT predictions do. For $T_\pi = 305 $ MeV, the model
reproduces the experimental cross sections above $M_{\pi\pi}^2 = 6 M_\pi^2$,
but overestimates the data below $M_{\pi\pi}^2 = 6 M_\pi^2$.
  We note again that the
 higher order terms generated by the unitarization procedure play only a
 small role at low energies.

% At $T_\pi= 190$ MeV,
%the $\Delta$-resonance can be excited. Since CHPT includes
%the dominant effects of the $\Delta$-resonance via the
%low-energy constants $c_2, c_3$, and $c_4$, it works well 
%even for relatively large energies of the incoming pion.

\begin{figure}
\resizebox{0.50\textwidth}{!}{
  \includegraphics{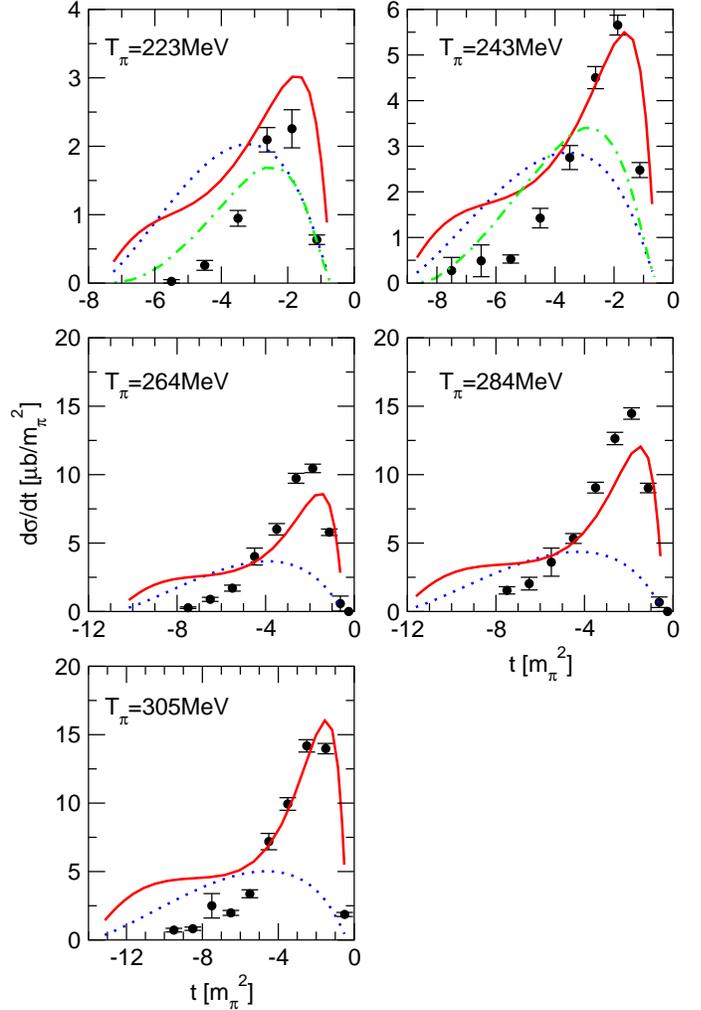}
}
\caption{Differential cross sections $d\sigma / dt$
 for  $  \pi^+ p \rightarrow \pi^+ \pi^+ n$.
Full model: (red) solid line; phase space: (blue) dotted line; 
chiral perturbation theory: (green) dot-dashed line~\cite{Fettes:1999wp}. 
 The data are from Ref.~\cite{Kermani:1998gp}.  
}
\label{fig:4}       % Give a unique label
\end{figure}

In contrast to the $M_{\pi\pi}$ distributions shown
in Fig.~\ref{fig:3}, the differential cross sections 
$d\sigma / dt$ strongly deviate from phase space, see  Fig.~\ref{fig:4}.
Chiral perturbation theory reproduces the shapes of the $t$-distributions
rather well, but misses the strong rise of the data at small momentum transfers.
The model can reproduce the $t$-distributions nearly quantitatively for 
small values of the momentum transfer $t$, but overpredicts the data at large
momentum transfers.

\begin{figure}
\resizebox{0.50\textwidth}{!}{
  \includegraphics{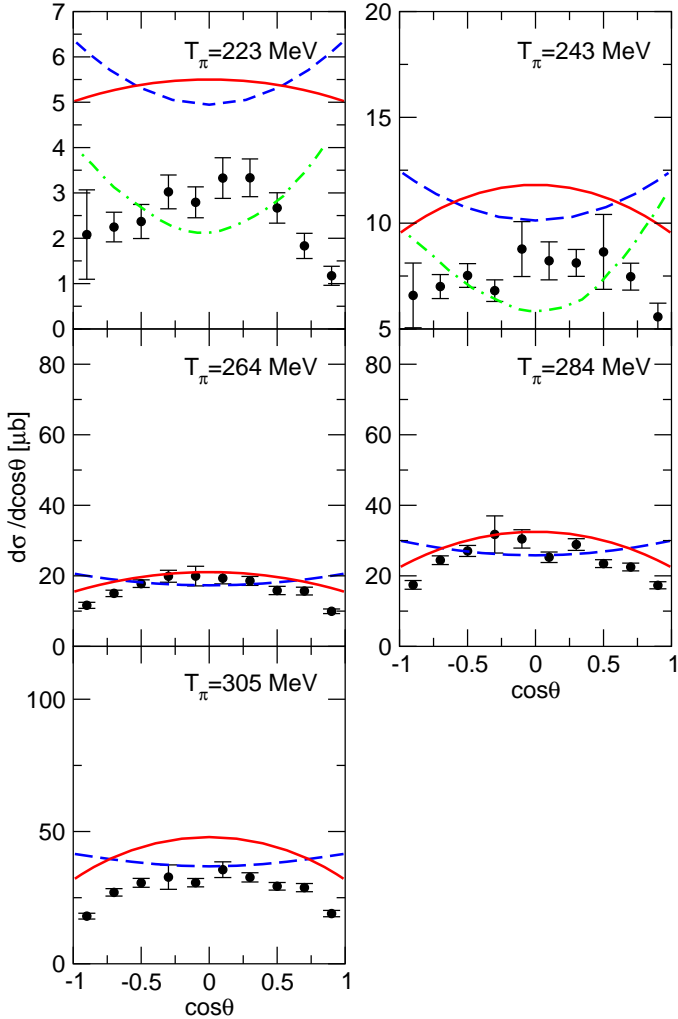}
}
\caption{Differential cross sections $d\sigma / d\cos\theta$
 for  $  \pi^+ p \rightarrow \pi^+ \pi^+ n$.
Full model: (red) solid line. A calculation employing bare
 vertices $f_0$ is displayed by the (blue) dashed line. The predictions of
chiral perturbation theory are shown by the (green) 
dot-dashed line~\cite{Fettes:1999wp}.
 The data are from \cite{Kermani:1998gp}.  
}
\label{fig:5}       % Give a unique label
\end{figure}

The differential cross section $d\sigma / d\cos\theta$ for the
$  \pi^+ p \rightarrow \pi^+ \pi^+ n$ reaction has to be symmetric
for $\cos\theta = 0$ because of the symmetry of the reaction under exchange of the
two produced pions. The experimental cross sections show a maximum at 
$\cos\theta = 0$, see   Fig.~\ref{fig:5}.
The chiral perturbation theory at order three predicts a minimum, however.
 The model obtains a maximum near $\theta = 90^\circ$
 in qualitative agreement with the data.
When replacing the dressed vertex functions $f$ by the bare ones, 
however, the shape of
the angular distribution changes and one obtains a minimum at
 $\theta = 90^\circ$.
This finding suggests that a chiral perturbation theory carried to fourth
order would solve the problem of the angular distributions in the 
$  \pi^+ p \rightarrow \pi^+ \pi^+ n$ reaction.

\begin{figure}
\resizebox{0.50\textwidth}{!}{
  \includegraphics{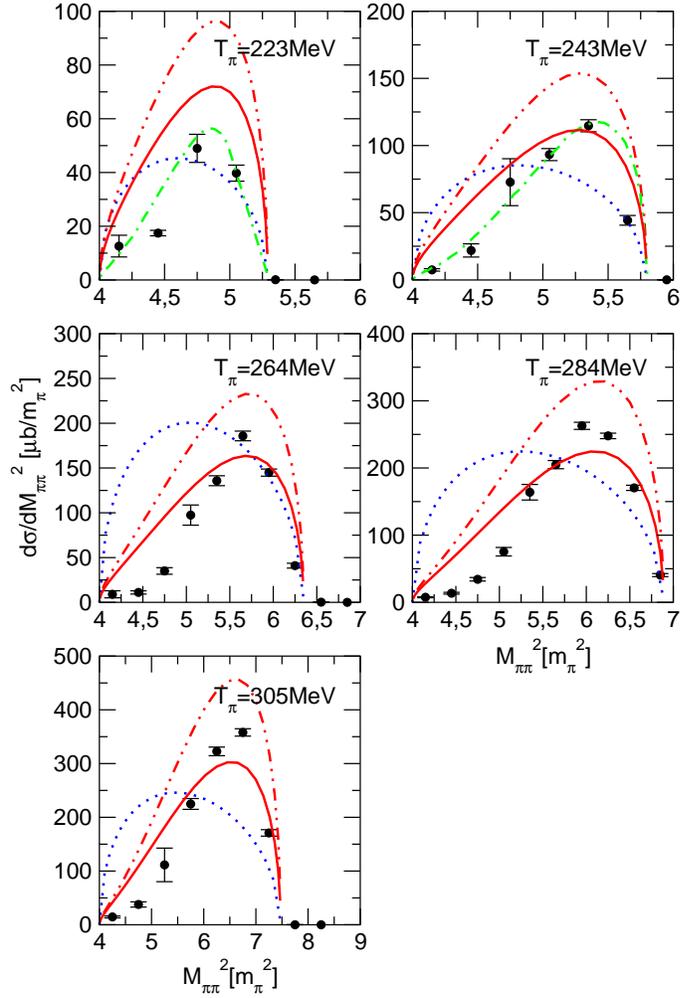}
}
\caption{Differential cross sections $d\sigma / dM^2_{\pi\pi}$
 for  $  \pi^- p \rightarrow \pi^+ \pi^- n$.
 Model without Roper resonance: (red) solid line;
 Model including Breit-Wigner Roper resonance: (red) double-dotted dashed line;
 phase space: (blue) dotted line; chiral perturbation theory: (green) dot-dashed 
 line~\cite{Fettes:1999wp}. 
 The data are from Ref.~\cite{Kermani:1998gp}.  
}
\label{fig:6}       % Give a unique label
\end{figure}

The differential cross sections ${d\sigma}/{dM_{\pi\pi}^2}$ for the reaction 
for  $  \pi^- p \rightarrow \pi^+ \pi^- n$ obtained at TRIUMF are shown in
Fig.~\ref{fig:6}.
The experimental data deviate significantly from phase space and show a pronounced
maximum close to the largest values of $M_{\pi\pi}^2$. Chiral perturbation
theory reproduces the asymmetric distributions very well for the pion kinetic
energies $T_\pi = 223 $ MeV and $T_\pi = 243$ MeV.
The model reproduces the shapes of the distributions, but overestimates the
absolute magnitudes. When switching off the contribution from the Roper resonance,
however, a reasonable reproduction of the data is achieved.
In the present model, we  have  treated the Roper in a simplified way,
 which apparently 
produces an overestimation of the effect of that resonance.
If one wants to incorporate the possibility to generate resonances dynamically,
a major revision of the present approach is required. 
In order to include the full non-pole $T$-matrix, an iteration of the
non-pole scattering kernel in the two-pion production diagrams has to be 
performed for different Lorentz frames.
 This goes beyond the scope of
the tree-level like model discussed here.

\section{Summary}

We have presented a resonance model for the 
$\pi N \to \pi\pi N$ reaction which incorporates information from
pion--nucleon scattering -- these two processes are intimately connected
and should not be treated independently from each other. The model is able 
to reproduce the total cross sections for kinetic energies of the incident pions up to 
about 350 MeV for all five reaction channels.
This is a success of the present approach which is only 
partially shared in specific reaction channels by other resonance
models which do not include consistently  pion-nucleon scattering
and two-pion production.
 The agreement with the differential experimental cross sections is
good, though not perfect. The total cross sections have a smooth
dependence on the pion kinetic energy which does not allow to
deduce obvious information about resonances.  The differential cross sections, on the
other hand,  show
characteristic deviations from  three-body phase space. The inclusion of 
the $\Delta$ resonance suffices to explain the overall shapes.

In the second resonance region, the model starts to break down. Here, we could
trace the difficulties to our approximate treatment of the Roper resonance.
The results obtained in the present study suggest to go beyond the Born
 approximation of  the non-pole pion-nucleon $T$-matrices, when constructing
the two-pion production amplitudes for analyses in the second resonance region.

\section*{Acknowledgements} 
We thank Christoph Hanhart for discussions.  
Partial support from  the EU Integrated Infrastructure
Initiative Had{\-}ron Physics Project (contract no. RII3-CT-2004-506078)
and DFG (SFB/TR 16, ``Subnuclear Structure of Matter'') is gratefully
acknowledged.

%\bibitem{cl56}  G. F. Chew and F. E. Low, 
%{\it  Phys. Rev.} {\bf 101}, 1570(1956).
%\bibitem{craig}  K. Craig  et al.,
%{\it  Phys. Rev. Lett.} {\bf 91}, 102301(2003).
\end{document}